\documentclass[review]{elsarticle}
\usepackage{lineno,hyperref}
\usepackage{graphicx}
\usepackage{color}
\modulolinenumbers[5]

\journal{Physics Letters A}

\begin{document}

\begin{frontmatter}

\title{A mechanical model of bacteriophage DNA ejection}

\author{Rahul Arun\fnref{myfootnote}}
\address{Adlai E. Stevenson High School, 1 Stevenson Drive, Lincolnshire, IL 60069}
\fntext[myfootnote]{presently at: California Institute of Technology, Pasadena, CA}
\author{Sandip Ghosal\corref{caSGH}\fnref{sghPhone}} 
\cortext[caSGH]{Corresponding author}
\fntext[sghPhone]{phone: (847)467-5990 fax: (847)491-3915 (Sandip Ghosal)}
\ead{s-ghosal@u.northwestern.edu}
\address{Department of Mechanical Engineering and Engineering Sciences and Applied Mathematics, Northwestern University, Evanston, IL 60208\\[4ex]
\copyright 2017 This manuscript version is made available under the CC-BY-NC-ND 4.0 license http://creativecommons.org/licenses/by-nc-nd/4.0/  \\[2ex]
DOI: 10.1016/j.physleta.2017.05.044}



\begin{abstract}
Single molecule experiments on bacteriophages show an exponential scaling 
for the dependence of mobility  on the length of DNA within the capsid.
It has been suggested that this could be due to the ``capstan mechanism'' -- the 
exponential amplification of friction forces that 
result when a rope is wound around a cylinder as in a ship's capstan. Here we 
describe a desktop experiment that illustrates the effect. 
Though our model phage is a million times larger, it exhibits the same scaling 
observed in single molecule experiments.
\end{abstract}


\end{frontmatter}

\linenumbers

\section{Introduction}
\label{sec:intro}
A bacteriophage is a virus that infects bacterial cells. Like all viruses, they lack the machinery to express the genetic information that they contain. Once inside their hosts, they ``hijack'' the cell's gene transcription mechanism to replicate themselves. Phages are essentially made of two components, a protein capsid
and the DNA (or RNA in case of RNA phages) that it encloses \cite{knobler_physical_2009}. 
Infection is initiated by the phage attaching to the cell 
membrane followed by injection of the DNA into the cell. The capsid remains outside attached to the cell
membrane. 
 
There are several known mechanisms that phages use to inject genetic material 
into  hosts~\cite{molineux_fifty-three_2006,inamdar_dynamics_2006,grayson_is_2007,molineux_popping_2013}. 
 In most double 
stranded DNA (dsDNA) phages, a fast ejection on the timescale of seconds can be
achieved by the release of elastic and electrostatic energy of the coiled up DNA confined within 
the viral capsid \cite{de_frutos_relationship_2013}. This is plausible because dsDNA has a persistence length $\sim 50$ nm and a 
very long strand (e.g. 48.5 kilobase pair or $\sim 17 \mu$m for the $\lambda$ phage)  
is packed into a capsid of diameter of the order of the DNA persistence length. In fact, for the $\lambda$-phage, the internal 
pressure in the phage capsid is estimated to be $\sim 50-100$ atmospheres \cite{de_frutos_relationship_2013}. It is nevertheless unlikely that 
this mechanism alone can explain DNA injection in all dsDNA phages. First, the driving pressure 
decreases rapidly as DNA empties the capsid; second, the interior of bacterial cells have 
a relatively high osmotic pressure $\sim 25$ atmospheres that would significantly retard  DNA 
entry. In some phages, the DNA ejection takes place in two steps. Initially, a part of the DNA 
is injected by this ``coiled spring'' mechanism. Subsequently, the inserted DNA is expressed to 
sythesize molecular motors which then reel in the remainder of the DNA by an ATP driven pulling 
action \cite{gonzalezhuici_pushpull_2004}. 

In order to understand the ejection process without too many layers of complexity, 
{\it in vitro} experiments have been designed \cite{grayson_real-time_2007} where $\lambda$ phages are induced to 
eject their DNA into the surrounding buffer in the absence of any host cells.
Here, it is indeed the elastic and electrostatic energy of the tightly coiled DNA within the capsid 
that drives the ejection. Furthermore, in the case of the $\lambda$ phage, the ejection 
takes place as a single continuous process without pauses and stops.
The driving force can be calculated from first principles by regarding the 
DNA as a charged semi-flexible rod \cite{purohit_mechanics_2003,purohit_forces_2005}.
The results of such calculations have been confirmed by 
experiments where the DNA is ``stalled'' after partial ejection by raising the osmotic 
pressure in the bath \cite{evilevitch_osmotic_2003,evilevitch_measuring_2004}. The speed of 
ejection, however, is determined not just by the 
stored potential energy in the capsid but also by the mechanisms of dissipation in the system, 
and this is more difficult to calculate.
Dissipation also plays a role in the opposite process of ATP driven packaging of 
DNA into the capsid. In that situation, it determines the time scale for reaching thermodynamic 
equilibrium. If the packaging rate is fast relative to this time scale then the DNA can take up 
a more disordered configuration than the ordered one that corresponds to the free energy minimum. 
On subsequent release from confinement, it encounters a higher frictional resistance as can 
be demonstrated in experiments as well as 
numerical simulations \cite{berndsen_nonequilibrium_2014,comolli_three-dimensional_2008}. 

It has been proposed \cite{ghosal_capstan_2012} that the dissipation arises from frictional forces 
between the DNA and the capsid wall and between neigboring 
strands as the helically wound DNA slides out of confinement. The frictional interaction is modeled 
by the classical Coulomb-Amonton laws
 \cite{marone_laboratory-derived_1998,persson_sliding_2000,bowden_friction_2001}
while disregarding the precise microscopic mechanisms. This is a 
plausible assumption as the Coulomb-Amonton
laws have been found to hold for nanoscale systems, though the underlying mechanism is quite different 
from that of the classical picture involving interlocking asperities 
\cite{mo_friction_2009,gao_frictional_2004,gnecco_velocity_2000}. The Coulomb-Amonton
 laws when combined with the 
equilibrium equations of an elastic rod lead to the conclusion that the tension in the rod increases 
exponentially with the length of DNA confined in the capsid  \cite{ghosal_capstan_2012}. 
This exponential amplification of tension
is a well known fact in mechanics and is known as the ``capstan principle''.
It is the principle of operation of a winch and various other familiar engineering 
innovations. The name derives from the cylinder or ``capstan'' used since antiquity for mooring ships.
Another example of the exponential amplification of friction forces is the extraordinary holding 
power of phone books with interleaved pages \cite{alarcon_self-amplification_2016,dalnoki-veress_why_2016}.
When applied to the ejection of dsDNA from phages, the capstan principle 
leads to the conclusion that the ejection  velocity per unit driving force (the mobility) should decrease exponentially 
with the amount of DNA confined within the capsid at any given instant. This is consistent with 
{\it in vitro} experiments on $\lambda$-phages \cite{ghosal_capstan_2012}.

Since the capstan mechanism also operates on macroscopic scales, it should be possible to 
demonstrate the exponential decrease of mobility on confined length using a centimeter scale model 
of the phage. In this paper we describe a table top experiment performed using everyday objects
that show this dependance. The paper is organized as follows. The main theoretical ideas 
relating to the capstan effect in phages are summarized in section~\ref{sec:capstan}. 
In section~\ref{sec:methods} a desktop experiment illustrating the mechanics of DNA 
ejection from phages is described. In section~\ref{sec:results} the data from the experiment 
is analyzed in the light of the theoretical ideas discussed earlier in section~\ref{sec:capstan}.
The significance of the experiment in the context of the DNA-phage problem and its 
limitations are discussed in section~\ref{sec:discussion}. Finally, the main conclusions 
are summarized in section~\ref{sec:conclusion}.

\section{The Capstan Model} 
\label{sec:capstan}
\begin{figure}[t]
   \centering
   \includegraphics[width=2in]{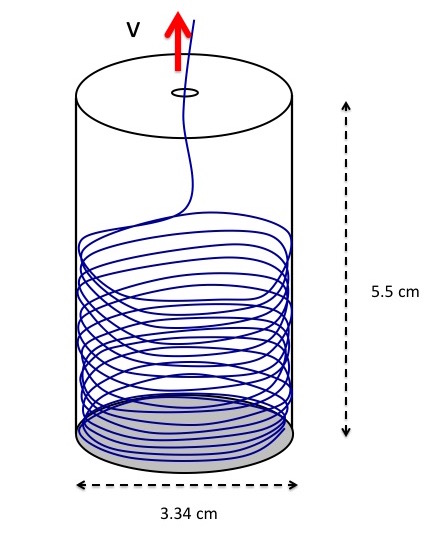} 
   \includegraphics[width=2in]{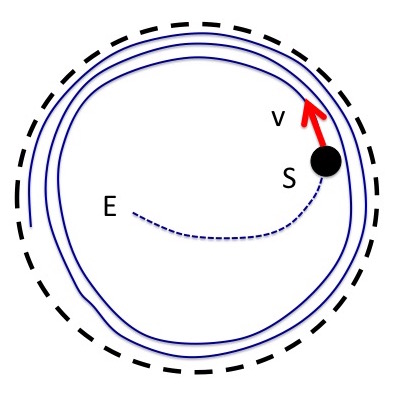} 
   \caption{{\it Left:} Schematic drawing of the mechanical model of a bacteriophage (see Supplementary Materials for 
   photograph and video of experiment). {\it Right:} Top view showing the spiral arrangement of the nylon 
   fiber within the container. 
   The fluid resistance on the trailing end SE is represented by the drag ($kv$) 
   on a single hypothetical particle indicated here 
   by the filled circle.}
   \label{fig:mechanical}
\end{figure}
Figure~\ref{fig:mechanical} is a schematic diagram of our experimental set up 
using a plastic container (representing the viral capsid) and a nylon filament (representing the DNA). 
The question that we wish to answer is the following: how 
does the length of fiber remaining within the capsid ($x$) vary with time ($t$) during the ejection 
process? 

The driving force $F$ in the ejection process can be obtained from an energy argument.
The driving force due to elasticity of the fiber is the same as the inward directed external force $F$ 
that must be applied to the DNA at the capsid exit to prevent it from exiting. The amount of this force 
can be calculated from the principle of virtual work 
\begin{equation} 
F \; dx = d {\cal F} 
\label{extF}
\end{equation} 
where ${\cal F} (x)$ is the free energy of the DNA. 
From the
theory of linear elasticity, the elastic bending energy of the DNA may be written as 
\begin{equation} 
{\cal F} (x) = \int_{0}^{x} \frac{EI}{2} \kappa^2 (s) \; ds
\label{elastic_energy}
\end{equation} 
where $E$ is the Young's modulus of the material, $I$ is the area moment of inertia of 
the fiber about the neutral plane, and, 
$\kappa (s)$ is the local radius of curvature of the centerline at a distance `$s$' from the 
capsid exit. In our mechanical model, the volume of nylon fiber inserted is only a small fraction 
of the capsid volume so that $\kappa (s) \approx 1/R$, $R$ being the radius of the cylindrical 
container. Thus, under these low packing conditions,
\begin{equation} 
{\cal F} (x) \approx  \frac{EI}{2R^2} x,
\label{elastic_energy1}
\end{equation}
so that the driving force is 
\begin{equation} 
F = {\cal F}^{\prime} (x) = \frac{EI}{2R^2}.
\label{constantF}
\end{equation} 
In the absence of an external stalling force, a tension 
\begin{equation} 
T(0) = F = \frac{EI}{2R^2}
\label{T(0)}
\end{equation} 
acts on the DNA at the capsid entrance that must be balanced by frictional 
resistance on the filament arising from within the capsid.

If we assume that the fiber may be described by the equations of elastic equilibrium of a 
beam and frictional forces are governed by the Coulomb-Amonton law with coefficient of kinetic 
friction ($\mu_k$), then\footnote{Eq.~(13) of \cite{ghosal_capstan_2012} contains an error:
the factor of $2 \pi$ in the exponent is superfluous.} \cite{ghosal_capstan_2012}
\begin{equation} 
T(s) = T(0) \exp \left( \frac{\mu_k s}{R} \right) = \frac{EI}{2R^2} 
\exp \left( \frac{\mu_k s}{R} \right),
\label{T(s)} 
\end{equation} 
neglecting any reduction in coil radius due to high packing fractions.
Equation~(\ref{T(s)}) is the Euler-Eytelwein formula. It is
well known for the problem of a flexible or semi-flexible string wrapped around a cylinder 
\cite{stuart_capstan_1961}, but is 
also applicable, as in this case, to the ``inverted'' problem of a semi-flexible rod confined 
within a hollow cylinder \cite{ghosal_capstan_2012}.

The constant curvature configuration of the beam 
cannot extend to the extremity E. This is because the internal bending moment is proportional 
to the curvature and the internal bending moment must vanish at the free boundary E.
What this means is that the fiber will lose contact with neighboring strands 
at some intermediate point S (Figure~\ref{fig:mechanical}) before E. Determining the true configuration 
of the fiber under these circumstances then becomes a difficult free boundary problem
since neither the location of E nor the shape of the section SE is known apriori. An analogous problem 
where a beam is pushed onto a hard surface from a point a fixed distance above it 
has been analyzed recently \cite{sano_slip_2016} and is shown to exhibit hysteresis of shape 
controlled by the static friction coefficient. In Brownian molecular dynamic simulations of bead-chain models, 
helical arrangements are spontaneously generated except for the trailing ends of the chain \cite{kindt_dna_2001}.
In order to avoid the complexity  of having to solve free boundary problems, we 
introduce a ``lumped parameter'' model for the resistance (possibly arising out of a combination 
of frictional and hydrodynamic forces) on the trailing end of the 
DNA that is not part of the helical arrangement. We do this by supposing that  the 
helically wound fiber is terminated
(point S in Figure~\ref{fig:mechanical}) by a bead that experiences a 
hydrodynamic drag $k v$ where 
$k > 0$ is a drag coefficient. Thus, 
\begin{equation} 
T(x) = \frac{EI}{2R^2} \exp \left( \frac{\mu_k x}{R} \right) = k v = - k \frac{dx}{dt}.
\end{equation} 
The solution of this differential equation then gives the time dependence of the length
$x$ remaining within the capsid 
\begin{equation} 
x = L - \frac{R}{\mu_k} \ln \left[ 1 +  \frac{E I \mu_k}{2 k R^3} \exp\left( \frac{\mu_k L}{R} \right) \; t \right] 
\label{x(t)}
\end{equation} 
where $L$ is the length of fiber initially in the capsid ignoring the short section SE. 
At short times, $t \rightarrow 0$, we have 
\begin{equation} 
x = L - v_m t 
\end{equation} 
where 
\begin{equation} 
v_m = \frac{E I}{2 k R^2} \exp\left( \frac{\mu_k L}{R} \right).
\label{vm}
\end{equation}
Eq.~(\ref{x(t)}) may be rewritten as 
\begin{equation} 
x = L - \frac{R}{\mu_k} \ln \left( 1 + \frac{\mu_k}{R} v_m \; t \right)
\label{x_norm}
\end{equation} 
and on differentiation,
\begin{equation} 
v = - \frac{dx}{dt} = \frac{v_m}{1 + t / \tau }
\end{equation} 
where 
\begin{equation} 
\tau^{-1} =  \mu_k v_m / R.
\label{tau}
\end{equation} 
Thus, the velocity decreases monotonically from a maximum of $v_m$
on a time scale $\tau$.
The total time of ejection $T$ is obtained by setting $t=T$ and $x=0$ in 
equation (\ref{x_norm}) 
\begin{equation} 
\frac{L}{v_m \tau} =  \ln \left(1 + \frac{T}{\tau} \right).
\label{scaling}
\end{equation} 
Here we have assumed that the length of the residual portion of the fiber that 
is not ejected may be taken as equal to the length SE in Figure~\ref{fig:mechanical}.

\section{Methods}
\label{sec:methods}
The bacteriophage model consists of a cylindrical (height, h=5.5 cm, radius, R=1.67 cm) 
polyethylene film canister (Fuji 35 mm, type: frosted clear, material: HDPE body, 
LDPE cap) representing the capsid and a nylon fishing line 
(Berkley Trilene Big Cat Monofilament Line, 0.55 mm average diameter, 13.6 kg break strength)
representing the viral DNA. A centered 2 mm diameter hole was drilled into the top of the film canister through which the monofilament was packaged and ejected. Pieces of filament were pre-cut to the following lengths for the experiment 
$L_{max} = 50, 100, 150, 200$ and $290$ cm. 
 The capsid was loaded by inserting one end of a filament into the container and slowly pushing in a pre-cut section of filament. As soon as the length of filament exceeded some multiple of the canister height, the filament spontaneously arranged itself in a helical pattern against the inside wall of the canister
  just as observed in Brownian molecular dynamic simulations of DNA-phage systems \cite{kindt_dna_2001}. 
 
 The filament was confined  by covering the exit 
hole with the thumb and released by removing the thumb. The event was recorded with a movie 
camera (Drift Innovation HD170 Action Camera, maximum frame rate of 60 fps and resolution 720p)
and the ejection time was measured with a stop watch (iPhone 5, Apple Inc.) 
measuring to $0.01$ s.
The time between the instant the hole was opened and the instant the filament stopped moving 
was recorded. In all cases, the ejection stopped with a variable amount of residual length remaining in the capsid. 
The recorded value of $L$ was taken as the actual length of filament pushed out of the capsid, that is, 
the true length ($L_{max}$) minus the residual length.
A similar behavior is also seen in the $\lambda$ phage {\it in vitro} experiment \cite{grayson_real-time_2007}
where there is a long pause between the time when the ejection stops and the time at which the DNA 
finally  separates from the capsid. The experiment was repeated several  times 
with each filament section
and  the ejected length $L$ and ejection time $T$ was recorded. This data is shown in 
Section 3, Table~1 of the accompanying Supplementary Material. It is seen that there is a spread in 
the measured values of $L$ and $T$ by an amount  $\Delta L \sim \pm 2$ cm  and $\Delta T \sim \pm 0.5$ s.
The experiment was then repeated by submersing the system first in filtered water 
(Ice Mountain 100 \% Natural Spring bottled drinking water) and then in glycerin 
(Soap Expressions 100 \% Vegetable Based). In each case, care was taken to ensure 
that the liquid completely filled the container without visible air bubbles. 
The fiber was pre-loaded prior to flooding 
the container with the fluid.
The corresponding 
ejected length ($L$) and ejection time ($T$) data are recorded in 
Section 3, Table~2 (water) and Table~3 (glycerin) of the accompanying Supplementary Material. 
The displayed data are an average over four separate runs (three, in the case of the longest 
fiber length) of the experiment.
A still image of the loaded capsid (Section 1) and a slow motion video of the capsid firing (Section 2) 
are also included.

\section{Results}
\label{sec:results}
\begin{figure}[t]
   \centering
   \includegraphics[width=4.5truein]{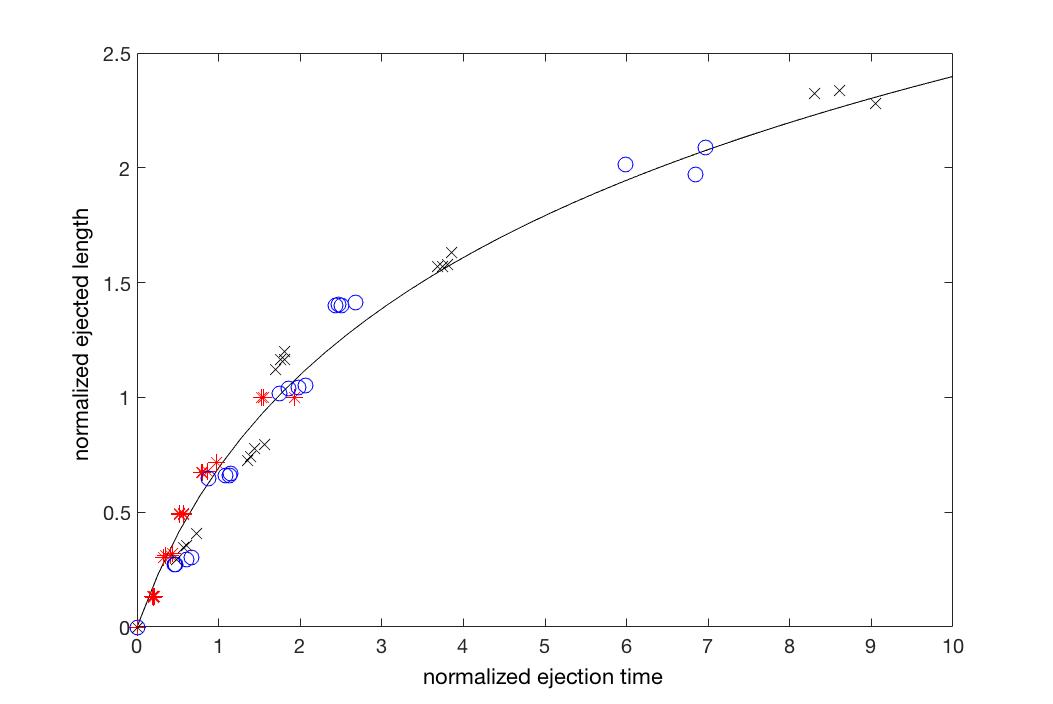} 
   \caption{The normalized ejected length $L/(v_m \tau)$ plotted against the normalized 
    ejection time $T/ \tau$ for air (star), water (circle) and glycerin (cross) ejections. The solid 
    line is equation~(\ref{scaling}).}
   \label{fig:scaling}
\end{figure}

The ejection process appeared qualitatively similar in all three media except for the speed of ejection.
In water, the process was slowed down by about a factor of five relative to that in air and by about a 
factor of almost two hundred in glycerin. Notwithstanding the widely different translocation times, equation 
(\ref{scaling}) provided a good fit to the data in all cases as shown in Figure~\ref{fig:scaling} using 
the similarity variables $L/(v_m \tau)$ and $T/\tau$. 
\begin{table}[b]
\centering
\begin{tabular}{| c | c | c | c | c |} \hline
 & Unit & Air & Water & Glycerin \\  \hline 
$v_m$ & cm/s & 140 & 70 & 1.6 \\
$\tau$ & s & 2.0 & 1.9 & 74 \\
$\mu_k$ & -- & 0.006 & 0.013 & 0.014 \\
\hline 
\end{tabular}
\caption{Scaling parameters $v_m$ and $\tau$ for air, water and glycerin ejections for the 
data shown in Figure~\ref{fig:scaling}. The corresponding friction coefficient $\mu_k$ is 
evaluated from equation~(\ref{tau}).}
\label{myTable}
\end{table}
For each data set, the two parameters $v_m \tau$ and $\tau$ were chosen to obtain the 
best fit to equation (\ref{scaling}). The fits were arrived at iteratively by trial and error 
and goodness of fit was decided visually. The best fit parameter values thus obtained are 
summarized in Table~\ref{myTable}. 
In each of the three cases, the value of the kinetic friction coefficient $\mu_k$ inferred from 
equation (\ref{tau}) is also noted.  In all three cases, notwithstanding the wide range of velocity 
and time scales, the friction coefficients are of similar magnitude. Also, the inferred value 
is not widely different from published values for lubricated polymer materials, though 
a detailed comparison will not be justified due to the rough nature of these experiments. 

\section{Discussion}
\label{sec:discussion}
The desktop experiment described here is only meant to illustrate the part 
of the DNA-phage system behavior relating to the capstan mechanism and is not necessarily an 
appropriate model for all aspects of the DNA ejection problem in phages. Indeed the two systems 
differ widely in spatial scales as well as in the details of the physical interactions that control 
their behavior.

An important difference between the current experiment 
and the real DNA-phage system is the extent of packing of the capsids. In $\lambda$-phage,
the volume of the polymer relative to the maximum available capsid volume (the packing 
fraction, $f$) is in the range $0.4-0.5$ \cite{grayson_effect_2006}. In the mechanical 
model presented here,  the packing fraction is easily estimated from the 
geometric parameters in section~\ref{sec:methods}, $f \approx 0.002-0.015$, much smaller 
than the actual values encountered in phages.  The behavior at high packing fractions 
has been explored recently using macroscopic mechanical models to represent 
the phage-DNA system 
\cite{gomes_geometric_2008,sobral_unpacking_2015,de_holanda_scaling_2016}. The focus of these investigations is the shape acquired by the packed DNA inside the capsid rather than the dynamics of ejection studied here.
Also, the metal wires used have significantly higher plasticity 
than the nylon fibers used in the present experiments.

The low packing fractions used here is also responsible for an important qualitative difference 
between our model and the real phage system.
Measured ejection velocities in the $\lambda$ phage when plotted as a function of the DNA remaining within the 
capsid show a characteristic unimodal shape. That is, the escape velocity first increase as the capsid starts to 
empty, reaches a maximum at an intermediate stage and finally decreases. This is because the velocity 
 $v = F(x) m(x)$ where the mobility $m(x) \propto exp(-bx)$ ($b > 0$) but the driving force $F$ is an 
 increasing function of $x$ since the capsid pressure increases as the capsid is tightly packed. In fact, when 
 the steric constraint is accounted for, $F(x) \propto (1-x/x_m)^{-1}$ where $x_m$ is the maximum length of 
 DNA that the capsid can accommodate \cite{purohit_mechanics_2003,purohit_forces_2005}.
 In the experiment described here, 
 the volume fraction of fiber inserted  was small compared to the maximum capacity of the capsid, 
 to avoid jamming and other practical complications. Thus,
 $F(x)$ is a constant given by equation (\ref{constantF}) so that the velocity decreases 
 monotonically rather than show a peak at an intermediate time.
 
The scale of the DNA-phage problem is such that any hydrodynamic interactions would be firmly 
in the Stokesian regime, that is, the Reynold's number $Re = \rho v d / \mu$, where $\rho$ is the fluid 
density, $\mu$ is the viscosity, $v$ is a characteristic velocity and $d$ is a characteristic length, 
is essentially zero.  In our mechanical 
model, using the exit hole diameter ($2$ mm) as a length scale and the maximumum value of the 
ejection velocity $v_m$ as the velocity scale, we find $Re = 374$, $1$ and $0.02$ respectively 
for ejections in air, water and glycerin. Thus, only the glycerin experiments truly 
represent the hydrodynamic
conditions in the DNA-phage system. However, we find that the scaling law, equation (\ref{scaling}), holds 
irrespective of the working fluid though the ejection velocity itself depends on fluid viscosity. 
This is consistent with the capstan model that attributes
the exponential dependence of mobility on filament length to interstrand friction, and 
fluid viscosity only plays an incidental role \cite{ghosal_capstan_2012}.

The pressure within the viral capsid ejecting the DNA arises from a combination of the bending 
energy of the DNA and the electrostatic self-repulsion between strands. These two factors contribute 
about equally to the potential energy of the confined DNA \cite{de_frutos_relationship_2013}. The electrostatic 
part of the potential energy can be readily tuned by changing the ionic composition of the solvent and 
thereby changing the Debye length. At high ionic concentrations and in the presence of polyvalent ions 
such as Mg$^{2+}$ the DNA backbone charge is almost perfectly shielded, reducing the electrostatic 
self-repulsion to almost zero. Ionic composition changes are found to affect the ejection speed 
in experiments \cite{grayson_real-time_2007} mainly through its effect on the capsid pressure. 
Numerical simulations suggest \cite{i_influence_2011} that the topology of the DNA configuration 
during the packaging phase can depend on the ionic composition of the bath. When the packaged 
DNA is allowed to eject freely, the ejection speed could in turn depend on the topology of the 
packaged strand \cite{marenduzzo_topological_2013}. These aspects of the problem of course 
cannot be accessed using our mechanical model as the stored potential energy in our nylon fiber 
is purely elastic; unlike DNA, it does not have an electrostatic component.

\section{Conclusion} 
\label{sec:conclusion}
Recent advances in single molecule observation and manipulation have made it possible to 
study the packaging \cite{smith_bacteriophage_2001}
and ejection \cite{grayson_real-time_2007} of DNA from phages with {\it in vitro} techniques. Such 
experiments have resulted in quantitative data that require explanations and interpretations. 
These problems belong to the ``mesoscale'' where the familiar classical mechanics of solids 
and fluids may still be applied but often in association with effects such as statistical fluctuations 
and Debye layer physics that are peculiar to small scale systems.
 In this paper, we consider one such problem,
bridging a gap of almost seven orders of magnitude to take ideas very familiar in marine engineering (winches 
and capstans) and applying them in the realm of cell biology (phages and DNA).

The specific single molecule experiment of interest here involves the ejection of DNA by the
$\lambda$ phage into the surrounding buffer. It was shown that the experimental 
data on ejection velocity 
is consistent with a frictional resistance law that has an exponential dependence on the length 
of DNA in the capsid. A possible mechanism for this was suggested \cite{ghosal_capstan_2012} based on 
the ``capstan effect'', where the helical geometry of the DNA together with the Coulomb-Amonton friction law results 
in an exponential amplification of friction with increasing turns.

The conditions required for the capstan mechanism can be simulated in a large scale system just as 
easily as in the bacteriophage. In this paper we have presented results from a desktop experiment 
that mimicks the process of DNA ejection from bacteriophages. The measured ejection time as a 
function of ejected length was found to be consistent with a scaling law derived on the basis of the 
capstan model. The scaling law was found to be independent of the 
Reynold's numbers in the  range $\sim 0.01 - 300$. 
 
\paragraph{This research did not receive any specific grant from funding agencies in the public, commercial, or not-for-profit sectors}
\section*{Acknowledgement} We wish to thank Prof. Sigurdur Thoroddsen (King Abdullah University of Science 
and Technology) for helpful suggestions and the Department of Mechanical Engineering, Northwestern University
for loan of the video camera.
\section*{References}

\newpage
\section*{List of supplementary material}
\noindent The following supplementary material accompanies this paper 
\begin{enumerate}
\item Photograph illustrating experimental set up.
\item Video of experiment in progress.
\item Tabular data of measured ejection times.
\end{enumerate}
\end{document}